\documentclass[conference,a4paper]{IEEEtran}

\usepackage[utf8]{inputenc} 
\usepackage[T1]{fontenc}
\usepackage{url}              
\usepackage{cite}             

\usepackage[cmex10]{amsmath}  
\usepackage{amssymb}
\usepackage{braket}
\usepackage{algorithm}
\usepackage{algpseudocode}
\usepackage{comment}

\newcommand{\BP}{\operatorname{BP}}

\algnewcommand{\IfThenElse}[3]{
  \State \algorithmicif\ #1\ \algorithmicthen\ #2\ \algorithmicelse\ #3}
  
\usepackage{ulem}

\usepackage{mleftright}       
\mleftright                   

\usepackage{graphicx}         
\usepackage[space]{grffile}			
\usepackage{booktabs}         

\let\emph\textit
\usepackage{amsthm}
\theoremstyle{definition}		

\newtheorem{definition}{Definition}

\newtheorem{remark}{Remark}

\newcommand{\cB}{{\cal B}}
\newcommand{\cC}{{\cal C}}
\newcommand{\cG}{{\cal G}}

\newcommand{\cO}{{\cal O}}
\newcommand{\cS}{{\cal S}}
\newcommand{\mOSD}{\text{mOSD}$_4$\xspace}	\usepackage{xspace}

\usepackage[colorinlistoftodos,prependcaption]{todonotes}
\usepackage{xcolor}
\usepackage{xargs}

\newcommandx{\yellownote}[2][1=]{\todo[inline,linecolor=yellow,backgroundcolor=yellow!25,bordercolor=yellow,#1]{#2}}

\begin{document}

\title{On Belief Propagation Decoding of Quantum Codes with Quaternary Reliability Statistics} 

\author{\IEEEauthorblockN{Ching-Feng Kung,\, Kao-Yueh~Kuo, \,and\, Ching-Yi~Lai}
\IEEEauthorblockA{\textit{Institute of Communications Engineering,} 
\textit{National Yang Ming Chiao Tung University,} 
Hsinchu 300093, Taiwan. \\
 \{kykuo,   cylai\}@nycu.edu.tw}
}

\maketitle

\begin{abstract}

In this paper, we investigate the use of quaternary reliability statistics for ordered statistics decoding (OSD) of quantum codes. OSD can be used to improve the performance of belief propagation (BP) decoding when it fails to correct the error syndrome. We propose an approach that leverages quaternary reliability information and the hard-decision history output by BP to perform reliability sorting for OSD. This approach improves upon previous methods that separately treat X and Z errors, by preserving the $X/Z$ correlations during the sorting step. Our simulations show that the refined BP with scalar messages and  the proposed OSD outperforms previous BP-OSD combinations. We achieve thresholds of roughly 17.5\%--17.7\% for toric, surface, and XZZX codes, and 15.42\% for hexagonal planar color codes.

\end{abstract}

\section{Introduction}
\label{sec:intro}

Quantum states are fragile and quantum error correction is needed in quantum information processing. 
Quantum stabilizer codes bear similarities to classical linear block codes \cite{Kit95,CS96,Ste96,GotPhD,CRSS98,NC00} 
and allow binary syndrome decoding.
Among them,  quantum low-density parity-check (LDPC) codes are preferred for practical issues.
In particular, they can be decoded by belief propagation (BP) 
\cite{MMM04,PC08,Wan+12,Bab+15,ROJ19,KL20,KL21a}
like classical  LDPC codes \cite{Gal62,MN95,Mac99,Tan81,Pea88,KFL01}.
A  quantum LDPC code of Calderbank--Shor--Steane (CSS)-type~\cite{CS96,Ste96} can be  decoded by the BP algorithm for binary codes (referred to as BP$_2$) if the $X$ and $Z$ errors are treated separately. One may also use a quaternary BP algorithm (referred to as BP$_4$) if $X$ and $Z$ errors are correlated.  

Usually the decoding performance of BP on a quantum code is compromised by the structure of quantum codes.
Several remedies to the BP algorithm are introduced for improvements \cite{PC08,Wan+12,Bab+15,ROJ19,KL20,KL21a}. However, for quantum codes with high  degeneracy, such as   topological codes, BP may still not work and other decoding algorithms are employed, including minimum-weight perfect matching (MWPM) \cite{Edm65,WFSH10} and other decoding strategies \cite{DP10,BSV14,DN17}. 

Recent advances in BP algorithms have demonstrated that the refined BP$_4$ with added memory effects, known as MBP$_4$, as well as its adaptive version, AMBP$_4$, are capable of handling several topological codes while remaining nearly linear time complexity \cite{KL22}.

 An alternative successful approach is to use ordered statistics decoding (OSD) \cite{FL95} to refine the messages obtained from BP \cite{PK19,RWBC20}, similar to its application in classical coding theory \cite{JN06,EM06}.
In the event that BP fails to decode a syndrome for a quantum code, one can utilize OSD to reconstruct the most probable error \cite{PK19,RWBC20}. However, it is important to note that this post-processing step incurs a cubic time complexity overhead, despite its effectiveness.

In this paper, we study OSD in  quantum coding theory.
Since OSD is only employed when BP fails, which has  a low probability of occurrence, it can be seen as a remedy to BP.
The sorting process is crucial in OSD \cite{FL95} and one unsatisfying aspect of the OSD development for quantum codes in the literature is that  it only considers the statistics  of $X$ and $Z$ errors  separately \cite{PK19,RWBC20}.
%
%
We propose a reliability sorting algorithm for quaternary statistics.
Moreover, our algorithm takes into account the hard-decision history obtained during BP iterations. Our simulations demonstrate that the hard-decision history is a crucial factor in determining the reliability order for OSD.  

For convenience, our OSD procedure and that in \cite{PK19,RWBC20} will be referred to as OSD$_4$ and  OSD$_2$, respectively. 
%
We will provide comparisons to show that OSD$_4$
greatly improves OSD$_2$ if $X$ and $Z$ errors are correlated. 
A remark is that the OSD$_2$ schemes in \cite{PK19,RWBC20} are designed for CSS codes, while our OSD$_4$ applies to general quantum codes.

We conduct simulations of (M)BP$_4$+OSD$_4$ on several quantum codes over depolarizing errors, including toric codes, surface codes, color codes, and non-CSS XZZX codes, as well as a highly-degenerate generalized hypergraph-product (GHP) code.
Our proposed scheme outperforms previous BP$_2$+OSD$_2$ or BP$_4$+OSD$_2$ schemes in the literature. We achieve thresholds of roughly 17.5\%--17.7\% for toric, surface, and XZZX codes, and a threshold of 15.42\% for the (6,6,6) color codes.
Comparisons of our results with existing threshold values are provided in Tables~\ref{tbl:thld} and~\ref{tbl:AvsO}. These results suggest that there is potential for further improving BP algorithms.

%


\section{Quantum stabilizer codes} \label{sec:stb}
In this section, we review the basic of stabilizer codes.

Consider the $n$-fold Pauli group 
$$\mathcal{G}_n \triangleq \left\{ cB_1\otimes\dots\otimes B_{n}: c\in \{\pm 1, \pm i\}, B_j\in \{I, X, Y, Z \} \right\},$$
where $I=[{1\atop 0} {0\atop 1}],$ $ X=[{0\atop 1} {1\atop 0}],$ $ Z=[{1\atop 0} {0\atop -1}],$ $ Y=iXZ$.
Every nonidentity Pauli operator in $\mathcal{G}_n$ has eigenvalues $\pm 1$. Any two Pauli operators in $\mathcal{G}_n$ either commute or anticommute with each other. 
A \emph{stabilizer group} $\mathcal{S}$ is an Abelian subgroup in $\mathcal{G}_n$ such that $-I^{\otimes n}\not\in \mathcal{S}$ \cite{NC00}.
Suppose that 
$\mathcal{S}$ is generated by $n-k$ independent generators.
Then $\mathcal{S}$ defines an  $[[n,k]]$ stabilizer code $\mathcal{C}(\mathcal{S})$ that encodes $k$ logical qubits into $n$ physical qubits:	
$$\mathcal{C}(\mathcal{S}) = \left\{ \ket{\psi}\in  \mathbb{C}^{2^n} : g\ket{\psi} = \ket{\psi} ~\forall\, g \in \mathcal{S} \right\}.$$
The elements in $\mathcal{S}$ are called \emph{stabilizers}.
%

We consider independent depolarizing errors with rate $\epsilon$ so that a qubit independently suffers each $X$, $Y$, or $Z$  error with probability $\epsilon/3$ and no error with probability $1-\epsilon$.

A Pauli error $E\in\cG_n$ can be detected by $\cC(\cS)$ through measurements if $E$ anticommutes with some of the stabilizers. The (binary) measurement outcomes $z\in\{0,1\}^m$ of ${m\ge n-k}$ stabilizers $\{S_i\}_{i=1}^m$ that generate $\cS$ will be called the \emph{error syndrome} of~$E$.
Thus we have the following decoding problem:  given an error syndrome~$z$, find the most probable $\hat E\in\cG_n$ such that $\hat E E \in\cS$ up to global phase. We say that an error estimate $\hat{E}$ is \emph{valid} if it matches the syndrome.

Without loss of generality, we assume that each stabilizer $S_i$ is of the form  $S_i=S_{i1} \otimes S_{i2} \otimes \dots\otimes S_{in}$, where $S_{ij} \in \{I, X, Y, Z\}$. 
Then we can study the decoding problem in the binary vector space~\cite{GotPhD,NC00}, using a mapping $\tau:$ 
    $$I \mapsto [0|0],~ X\mapsto [1|0],~ Z\mapsto [0|1],~ Y\mapsto [1|1].$$
A \emph{check matrix} of the stabilizer code $\cC(\cS)$
is a binary matrix $\tilde S = [\tilde S_{ij}] \in\{0,1\}^{m\times 2n},$
where $$[\tilde{S}_{ij}|\tilde{S}_{i(j+n)}]=\tau(S_{ij}).$$ 
A Pauli error  $E$ can also be represented by a binary vector 
	$\tilde{E}	= [\tilde{E}^X | \tilde{E}^Z] \in \{0, 1\}^{2n}$,
 where $\tilde{E}^X$ and $\tilde{E}^Z $ are the indicator vectors of $X$ and $Z$ components of $E$, respectively.
 Then the error syndrome of $E$ is 
 \begin{align}
 z = \tilde{E}\Lambda \tilde{S}^T 
   = \tilde{E} (\tilde{S}\Lambda)^T \in\{0,1\}^m, \label{eq:syndrome}
 \end{align}
 where 
 $\Lambda = \left[ {O_n \atop I_n} \bigg| {I_n \atop O_n} \right],$ 
and $O_n$ and $I_n$ are $n\times n$ zero and identity matrices, respectively.

Consequently, the decoding problem is to solve this   system of linear equations with  $2n$ binary variables $\tilde{E}_j$ that are most probable.
Note that $ \tilde{S}$ is of rank $n-k$ so
the degree of freedom of this binary system is $n+k$.


\subsection{Belief Propagation (BP) Decoding} \label{sec:BPs}

Given a check matrix $\tilde{S}$, a syndrome $z$, and depolarizing error rate $\epsilon$,
the initial error distribution $p_i$ for  qubit $i\in\{1,2,\dots,n\}$ is described by a vector $ ({1-\epsilon}, \frac{\epsilon}{3}, \frac{\epsilon}{3}, \frac{\epsilon}{3})$,
which initializes the (output) belief vector
$q_i = (q_i^I, q_i^X, q_i^Y, q_i^Z)$
(or its log-likelihood version~\cite{KL21a}). 
BP will iteratively update $\{q_i\}_{i=1}^n$ 
 and make a hard decision for $\hat{E}\in\{I,X,Y,Z\}^n$ to match $z$ at each iteration.
A maximum number of iterations $T$ will be chosen in advance.
If a valid error estimate is generated before $T$ iterations, it will be accepted as the solution.
Otherwise, BP will  terminate and  claim a failure.


\section{Order-statistic decoding (OSD) based on quaternary BP (BP$_4$)} \label{sec:BPOSD}

If BP fails to provide a valid error estimate after $T$ iterations for the binary system~(\ref{eq:syndrome}) 
for a given parity-check matrix $\tilde{S}$ and an error syndrome $z$, OSD will be utilized. The critical step in OSD is identifying and sorting the correct and reliable coordinates  to determine $n+k$ more reliable bits.
 Therefore, we need to define the notions of reliability order first.

 We remark that a quantum code can be seen as a quaternary additive code with binary error syndromes. 
Thus one may formulate a system of $n$ quaternary variables and $n-k$ additive constraints as well. However, such a system is not linear and OSD cannot be applied before converting to a system in \eqref{eq:syndrome}.

\subsection{Reliability order based on BP$_4$} \label{sec:reliab}

To utilize the quaternary distribution generated by BP$_4$ for each qubit error in OSD, 
we propose a method that employs the hard-decision history from all BP iterations and the output probabilities from the last BP iteration.

Suppose that BP fails to provide a valid error after  $T$ iterations. At this point, we have the output distribution from the last BP iteration $q_i = (q_i^I, q_i^X, q_i^Y, q_i^Z)$, $i = 1, 2, \dots, n$. Additionally, we have  
the  hard-decision results for all $T$ iterations, denoted by $W = [W_{ji}] \in \{I, X, Y, Z\}^{T\times n}$.

Our objective is to assign reliability for each qubit error  accordingly.
We define two types of measures in the following.
\begin{definition} \label{def:LastFor} 
Define a  (hard) reliability vector $\ell\in\mathbb{R}^n$, where $\ell_i$ denotes the number of iterations during which the final hard decision for the error at qubit $i$ remains the same. 
\end{definition}
\noindent One can check that
$\ell_i = x$   if and only if $W_{j,i} = W_{T,i}$ for all $T-x+1 \leq j \leq T$, and   $W_{T-x, i} \neq W_{T,i}$.

\begin{remark}
For implementation, the values of  $\{\ell_i\}_{i=1}^n$ can be updated  using only $W_{r-1}$ and $W_r$ at iteration $r$. This means that storing the entire hard-decision history $W$ is unnecessary.
\end{remark}

 \begin{definition}  
Define two  (soft) reliability functions
\begin{align*}
    \phi^X(i) &= \max\{ q_{i}^X+q_{i}^Y, q_{i}^I+q_{i}^Z\},\\
    \phi^Z(i) &= \max\{q_{i}^Z+q_{i}^Y, q_{i}^I+q_{i}^X\},
\end{align*}
for  $i\in\{1,2,\dots,n\}$.
\end{definition}
\noindent Note that $\{q_{i}^X+q_{i}^Y, q_{i}^I+q_{i}^Z\}$ is a binary distribution indicating whether an $X$ error occurs or not at the $i$-th qubit.

Now we define a reliability order for the bits in $\tilde{E}	= [\tilde{E}^X | \tilde{E}^Z] \in \{0, 1\}^{2n}$ based on the above hard and soft reliability functions from BP$_4$  as follows.
\begin{definition} \label{def:reliability order}
    Error bit $\tilde{E}_{i}^a$ is said to be more reliable than error bit $\tilde{E}_{j}^b$ if $\ell_i >  \ell_j$, or if $\phi^a(i) \geq   \phi^b(j)$ when $\ell_i = \ell_j$, for $1\leq i, j\leq n$ and $a, b \in \{X, Z\}$. 
\end{definition}

For comparison, we  define a reliability order that accounts for only  the soft reliability functions of BP$_4$.
\begin{definition} \label{def:reliability order2}
Error bit  $\tilde{E}_{i}^a$ is said to be more  reliable than error bit $\tilde{E}_{j}^b$ in the marginal distribution if  $\phi^a(i) \geq   \phi^b(j)$.
\end{definition} 
An OSD algorithm that follows the reliability order in  Def.~\ref{def:reliability order}  or Def.~\ref{def:reliability order2} is referred to as  OSD$_4$  or \mOSD, respectively.

Note that  \mOSD is  different from  the OSD$_2$ schemes  in \cite{PK19,RWBC20}, where the reliabilities of $\{\tilde{E}_{i}^X\}_{i=1}^n$ and $\{\tilde{E}_{j}^Z\}_{j=1}^n$ are separately sorted.

\subsection{OSD$_4$ and its extension OSD$_4$-$w$} \label{sec:OSDw}

Now we provide OSD algorithms according to the reliability orders defined in the previous subsection. 
Suppose that $\tilde{S}$ and $z$ are given, and $\tilde{E}$ is the hard-decision output from BP$_4$.
(Eq.~(\ref{eq:syndrome}) is unsatisfied.)
The steps of our OSD$_4$ are as follows.
\begin{enumerate} 
 
	\item 
 
 Sort the error bits  in ascending order of reliability  based on  Def.~\ref{def:reliability order} (or Def.~\ref{def:reliability order2}).
	Construct a corresponding column permutation function $\pi$ 
        and calculate $\pi(\tilde{E} )$ and $\pi(  \tilde{S} \Lambda )$.
 
	\item 
	Perform Gaussian elimination on $\pi( \tilde{S} \Lambda)$, and  if necessary, apply a column permutation function $\mu$ to ensure that the first $n-k$ columns of the resulting matrix  $\tilde S'$  are linearly independent.  Then we have
	    $$\tilde{S}' = \left[
	    \begin{array}{cc}
	    I_{n-k} & A \\
	    O       & O
	    \end{array}\right],$$
	where the last $m-(n-k)$ rows are all zeros.

\item  Apply the same row operations used in the Gaussian elimination  process to $z^T$ and let its transpose be $z'$.
 Remove the last $m-(n-k)$ entries of $z'$. 
 
 \item  Let $\tilde{E}'= \mu(\pi(\tilde{E}))$ and denote $\tilde{E}' = \left[\tilde{E}'_U ~~ \tilde{E}'_R \right]$,
   where $\tilde{E}'_U$~is the  unreliable part of  $n-k$ bits and 
   $\tilde{E}'_R$ is the reliable part of $n+k$ bits. 
  Generate the error estimate 
  ${ \Hat{E} \triangleq \left[z'+\tilde{E}_R'A^T ~~ \tilde{E}'_R \right] }$.
\item Return   $\pi^{-1}(\mu^{-1}(\Hat{E}))$.

\end{enumerate}

The above procedure is also denoted   OSD$_4$-0.
If the bits in the reliable part are correct, the output of  OSD$_4$-0 is the required solution. 

If some bits in the reliable part are incorrect, additional processing is necessary. 
We may flip up to $w$ bits in the reliable part and  generate the corresponding error estimate. 
This process is repeated for all the $\sum_{i=0}^w \binom{n+k}{i}$ possibilities and a valid error with minimum weight is selected as the output. 
This type of OSD algorithm is referred to as \mbox{order-$w$}~OSD$_4$ (or OSD$_4$-$w$ for short).
(Note that unless otherwise specified, the {\it weight} of a binary vector is defined as the   weight of the corresponding Pauli error.)
The procedure of OSD$_4$-$w$ is outlined in Algorithm~\ref{algo:OSDw}.

\begin{algorithm}[t] \caption{: OSD$_4$-$w$} \label{algo:OSDw}
    \textbf{Input}: Check matrix $\tilde{S}$, error syndrome $z$, initial error $\tilde{E}$, 
    BP output distributions $\{q_i\}_{i=1}^n$, reliability vector $\ell$.

    \textbf{Output}: A valid error of $2n$ bits with minimum weight. 

    {\bf Step}: 
    \begin{algorithmic}
 
    \State 
    $\pi \gets \operatorname{SortReliability}(  \ell, \{q_i\}_{i=1}^n)$. \Comment{1)}
    \State $([I~A], z', \mu) \gets \operatorname{GaussianElimin}(\pi(\tilde{S}\Lambda), z)$. \Comment{2), 3)}
    \State  $\tilde{E}'=\left[ \tilde{E}_U' ~~  \tilde{E}_R' \right] \gets \mu(\pi(\tilde{E}))$,  \Comment{ 4)}
    \State $\Hat{E} \gets \left[z'+\tilde{E}_R'A^T ~~  \tilde{E}_R' \right]$. 
    \For {$\tilde{E}_R''\in$\{vectors different from $\tilde{E}_R'$ for at most $w$ bits\}}
    \State $\tilde{E}'' \gets \left[z'+\tilde{E}_R''A^T ~~  \tilde{E}''_R \right]$. 
    \State \algorithmicif \ {$\tilde{E}''$ has smaller weight than $\hat E$} 
    \algorithmicthen \ $\Hat{E} \gets \tilde{E}''$.
    \EndFor
    \State \Return $\pi^{-1}(\mu^{-1}(\Hat{E}))$. \Comment{5)}
    \end{algorithmic}
\end{algorithm}

\subsection{Complexity of OSD$_4$} \label{sec:cmplxty}

The complexity of OSD$_4$-0 is $\mathcal{O}(n^3)$ dominated by the Gaussian elimination step. 
For OSD$_4$-$w$ with $w>0$, the total number of  possible flips  is $\sum_{j=1}^{w}\binom{n+k}{j}$, which is  $\cO((n+k)^w)$ or $\cO(n^w)$ if $k$ is negligible with respect to $n$.
In Algorithm~\ref{algo:OSDw}, it is crucial to note that when we flip from $\tilde{E}_R'$ to get $\tilde{E}_{R}''$ and attempt to calculate $z'+\tilde{E}_{R}''A^T$, it can be efficiently obtained from $z'+\tilde{E}_{R}'A^T$ using $\mathcal{O}(n)$ calculations.
Therefore, the \mbox{OSD$_4$-$w$} algorithm has complexity $\mathcal{O}(n^3+n^{w+1})$, which is $\cO(n^3)$ when $w\le 2$.

It is worth noting that the above OSD complexity is only required when BP fails to converge. BP has a nearly linear complexity in $n$ and has a high probability of convergence for many spare-graph quantum codes \cite{KL20,KL21a}.

\section{Simulation results} \label{sec:sim}

\begin{algorithm}[t] \caption{: BP$_4$+OSD$_4$-$w$} \label{algo:BPOSD}
    \textbf{Input}: Check matrix $\tilde{S}$, error syndrome $z$, maximum number of iterations $T$, initial error distributions $\{p_i\}_{i=1}^n$.

    \textbf{Output}: A valid error of $2n$ bits with minimum weight.

    \textbf{Initialization}:
    \begin{algorithmic}
        \State Let $\cB$ be the beliefs  propagating in BP, which will be initialized by $\{p_i\}_{i=1}^n$.
        \State Let $\ell$    be an all-one vector of length $n$.
        \State Let $W_0$ be a Pauli vector of $I^{\otimes n}$.
    \end{algorithmic}
    {\bf Step}: 
    \begin{algorithmic}
    \For {$j = 1$ to $T$}
    \State $(\{q_i\}_{i=1}^n, \cB) \gets \BP_4(\tilde{S}, z, \{p_i\}_{i=1}^n, \cB)$ for one iteration.
    \State ${W}_1 \gets  \operatorname{HardDecision}(\{q_i\}_{i=1}^n)$.
    \State \algorithmicif\ {${W}_1$ matches the syndrome $z$} \algorithmicthen \ \Return ${W}_1$.
    \State {$\ell \gets \operatorname{UpdateReliabilityVec}(W_0, W_1, \ell)$.}
        \Comment{Definition~\ref{def:LastFor}}
    \State {$W_0 \gets W_1$.}
    \EndFor
    \State \Return $\Hat{E} \gets \text{OSD$_4$-$w$}(\tilde{S}, z, W_1, q_i\text{'s}, \ell$). \Comment{Algorithm \ref{algo:OSDw}}
    \end{algorithmic}
\end{algorithm}

We simulate the performance of BP$_4$+OSD$_4$-$w$, which is outlined in Algorithm~\ref{algo:BPOSD},
on a GHP code and various 2D topological codes (including CSS and non-CSS topological codes \cite{Kit03,BM06,BM07,HFDM12,THD12,KDP11,ATBFB21}, as summarized in \cite[Table~I]{KL22isit}).

It is worth noting that the performance of BP$_4$+OSD$_4$ 
is similar when using both parallel and serial schedules. However, for consistency with the settings in \cite{KL22,KL22isit}, we consider the serial schedule in the following simulations for comparison purposes. We set the maximum number of BP iterations to $T=100$ for the GHP code and $T=60$ for topological codes, although a much smaller value of $T$ is usually sufficient.
For each data point in each plot, at least 100 logical error events are collected.

\begin{figure}
	\centering\includegraphics[scale = 0.6]{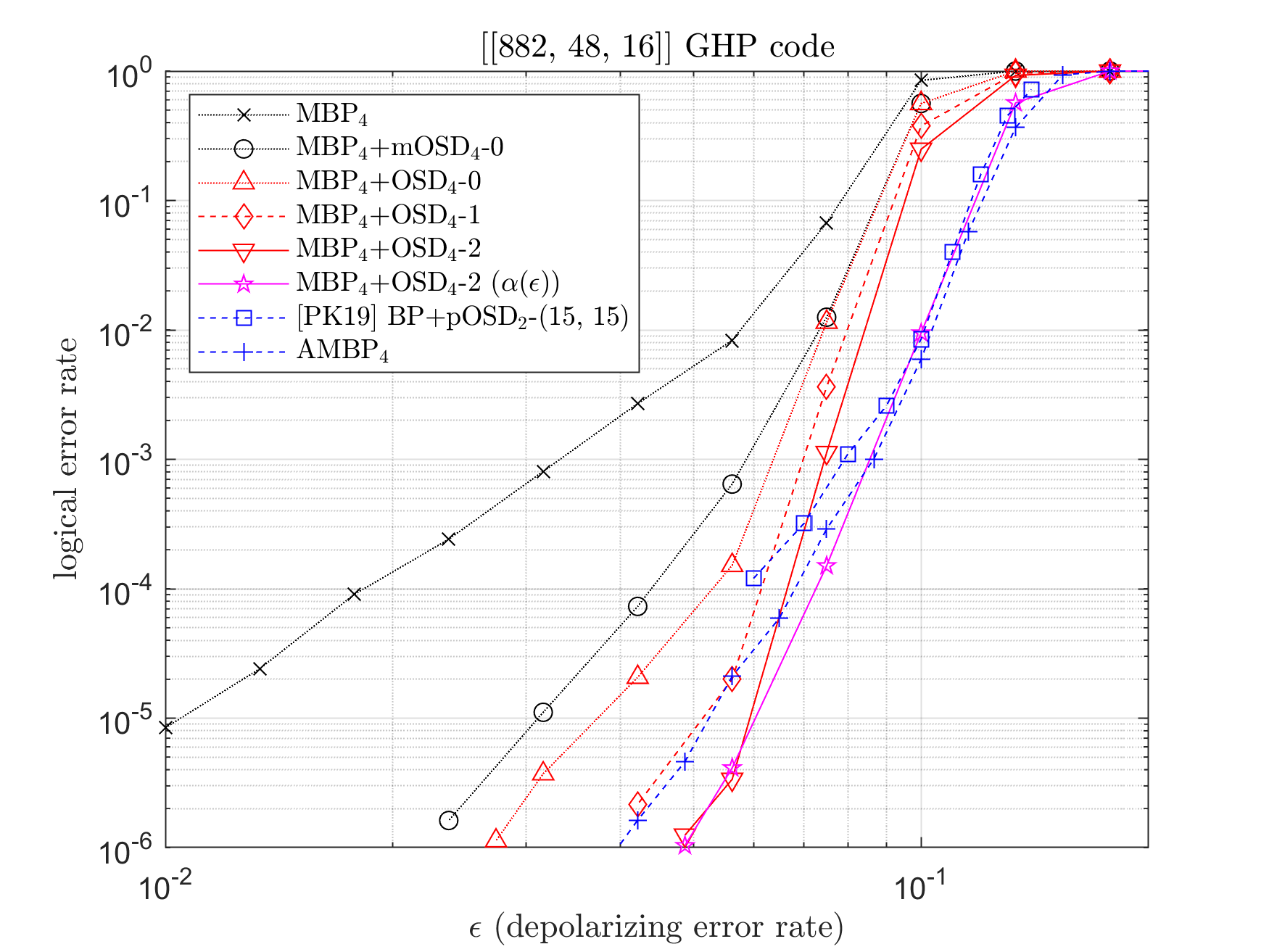}
	\caption{
		Comparison of various BP-OSD schemes on the $[[882,48]]$ GHP code. 
        The curve [PK19] BP$_4$+pOSD$_2$-(15,15) is from \cite{PK19}.
	}\label{fig:GHP}
\end{figure}

We begin by considering a highly-degenerate $[[882,48]]$ GHP code (with minimum distance $d=16$) provided in \cite{PK19} to compare various BP-OSD schemes.
In \cite{PK19}, they used a partial order-$w$ OSD that selects   $\sum_{j=0}^w\binom{\lambda}{j}$ candidates  for a parameter  $\lambda<n+k$,
and their decoding algorithm is denoted as BP$_4$+pOSD$_2$-$(w,\lambda)$.
We simulate the performance of different combinations of  MBP and OSD, as shown in Fig.~\ref{fig:GHP}. 
For reference, we also plot the performance of AMBP$_4$.
The results show that  OSD can significantly enhance the performance of MBP$_4$, with MBP$_4$+OSD$_4$-$2$ outperforming AMBP$_4$.

One can also observe that MBP$_4$+OSD$_4$-$0$ (with hard reliability) outperforms MBP$_4$+mOSD$_4$-$0$ (without hard reliability) by roughly half an order of performance.
This suggests that the hard reliability vector determined by the hard-decision history of BP is crucial in determining the reliability order for OSD.

Note that MBP$_4$ has a parameter $\alpha$ \cite{KL22}, which we set as fixed $\alpha=1.6$ or $\alpha(\epsilon)= -0.16 \log{10}(\epsilon) -0.48$, where the relation between $\alpha$ and $\epsilon$ is suggested in \cite[Fig.~3]{KL22}~(arXiv version), and the coefficients $-0.16$ and $-0.48$ are determined by pre-simulations.

\begin{figure}
	\centering \includegraphics[scale = 0.6]{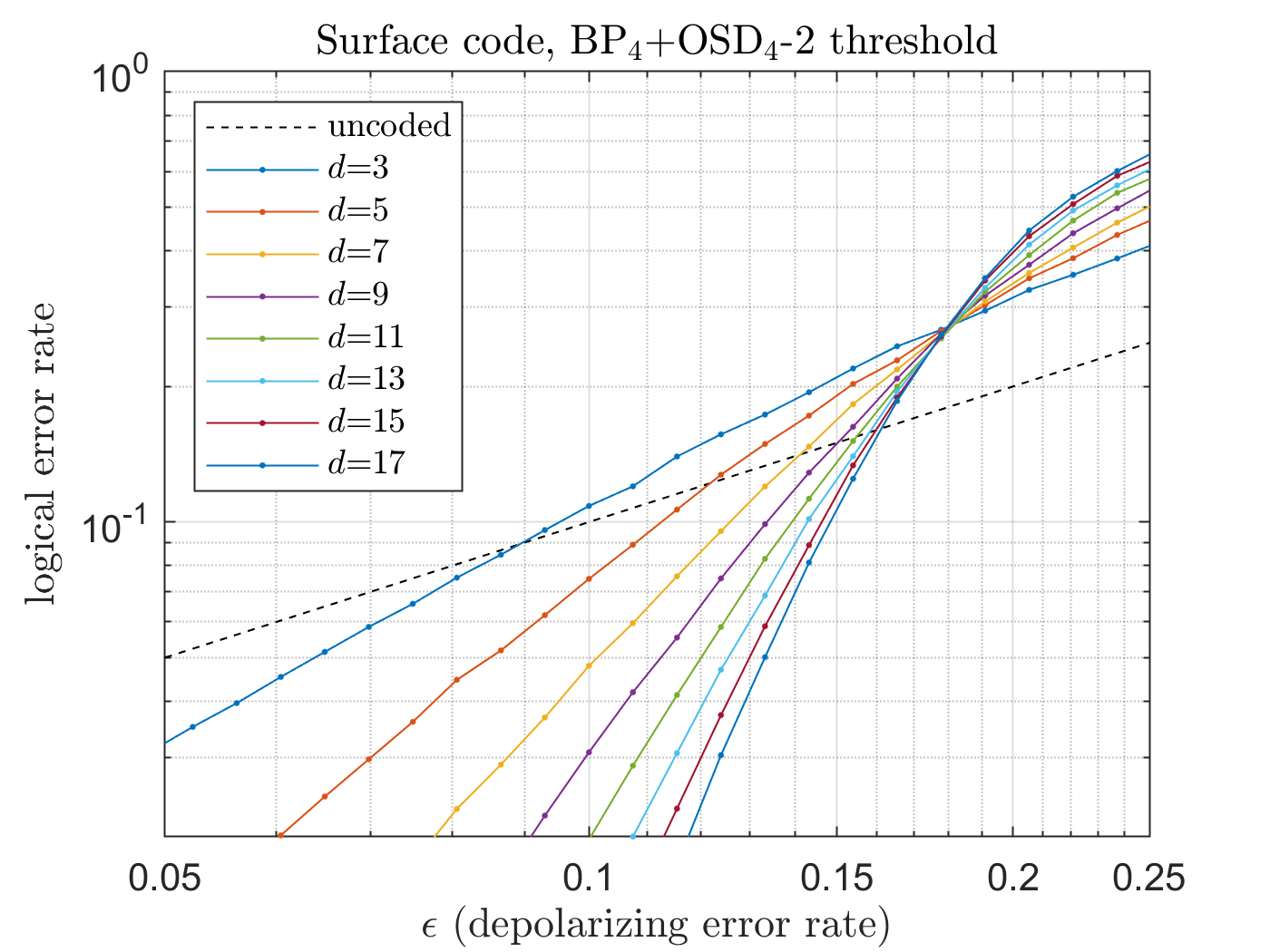}
	\caption{
		The threshold of BP$_4$+OSD$_4$-2 on the surface codes is about 17.68\% from the simulations. The dashed line stands for no error correction. 
        The used BP$_4$ can be considered as MBP$_4$ with $\alpha=1$. While using appropriate $\alpha(\epsilon)$ for MBP may result in better overall performance, this improvement may not be readily apparent in threshold analysis.
	}\label{fig:eps_surface} \vspace*{\floatsep}
	\centering \includegraphics[scale = 0.6]{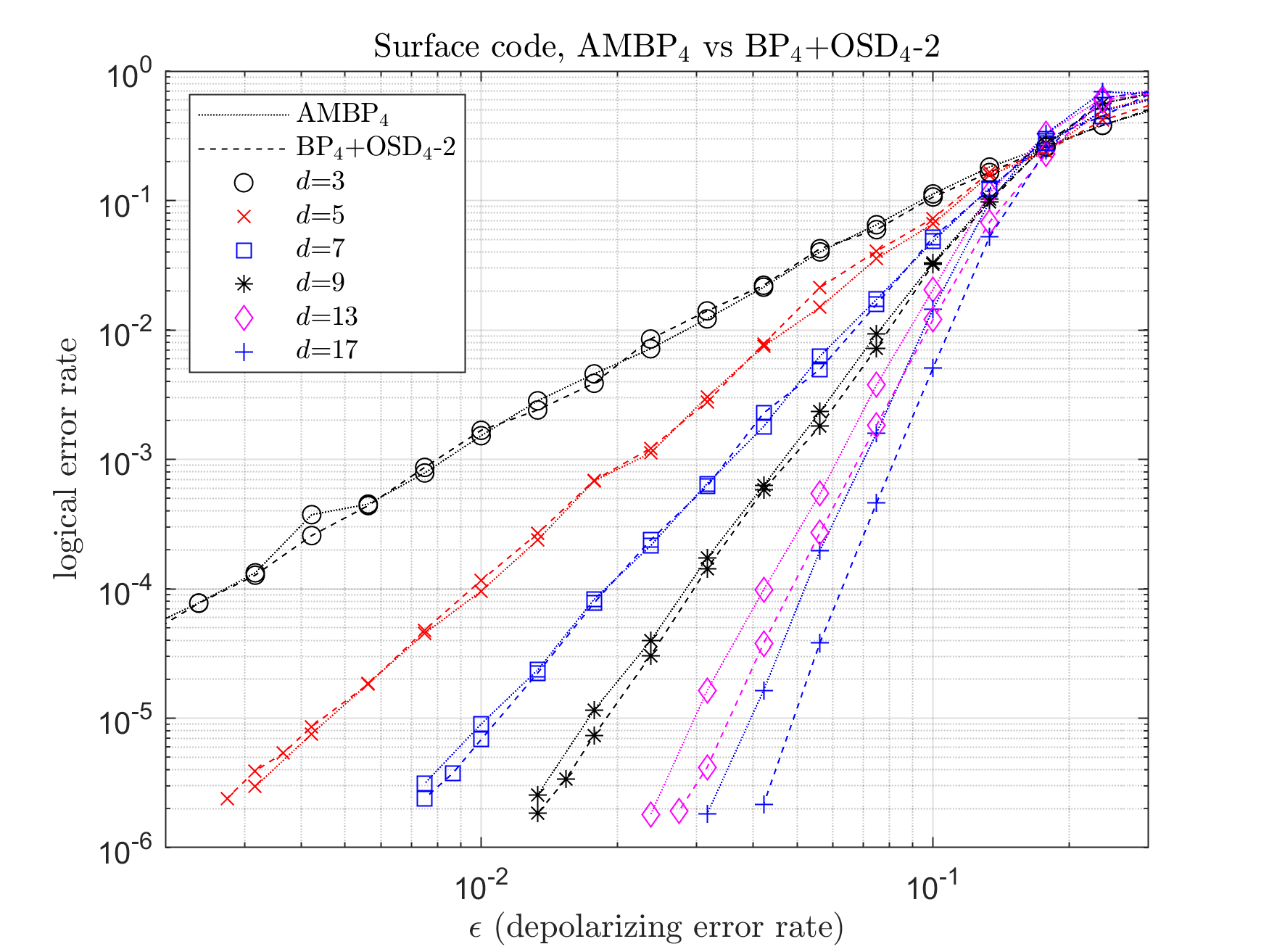}
	\caption{
		Comparison  of AMBP$_4$ and BP$_4$+OSD$_4$-2 on the surface codes.
	}\label{fig:surface}
\end{figure}

\begin{table}
	\caption{Thresholds of several decoders over depolarizing errors.} \label{tbl:thld} 
	\centering{
		$\begin{array}{|l|l|}
		\hline
		\text{decoder} 								& \text{threshold} \\
		\hline                                              
		\text{BP$_4$+pOSD$_2$-(15,15) \cite{PK19}}			& 15\%\text{ (GB codes)$^\dag$} \\
		\text{BP$_2$+pOSD$_2$-(2,60) \cite{RWBC20}}	    & 14.85\%\text{ (toric codes)$^\star$} \\
		\text{MWPM \cite{Edm65,WFSH10}}				& 15.5\%\text{ (toric codes)} \\
		\text{AMBP$_4$ \cite{KL22}}  				& 17.5\%\text{ (toric codes)} \\
		\text{BP$_4$+OSD$_4$-2 (this paper)}		& 17.52\%\text{ (toric codes)} \\
		\hline
		\end{array}$}
	\\
	\begin{flushleft}
		$^\dag$:  \cite{PK19} did not simulate topological codes, but instead focused on generalized bicycle (GB) codes.
        \\
        $^\star$: this is estimated from the threshold of 9.9\% in the bit-flip channel.
	\end{flushleft} \vspace*{\floatsep}
%
%
	\caption{Thresholds of AMBP$_4$ and BP$_4$+OSD$_4$-2 on several topological code families over depolarizing errors.} \label{tbl:AvsO} 
	\centering{
		$\begin{array}{|l|l|l|l|l|}
		\hline
		\text{decoder} 						& \text{toric}	& \text{surface}	& \text{color}   & \text{XZZX}\\
		\hline                                              
		\text{AMBP$_4$ \cite{KL22,KL22isit}}& 17.5\% 		& 16\%				& 14.5\%		& 17.5\% \\
		\text{BP$_4$+OSD$_4$-2 (this paper)}& 17.52\% 		& 17.68\%			& 15.42\%		& 17.72\% \\
		\hline
		\end{array}$
		}
		\\[3pt]
	\begin{flushleft}
        %
        The color codes are the planar (6,6,6) color codes, and the XZZX codes have a twisted structure on the torus \cite{KL22isit}.
	\end{flushleft}
\end{table}

Next, we consider 2D topological codes, whose decoding performance is typically evaluated by the tolerable error threshold. See a comparison in \cite[Table~II]{KL22isit}.
 We use BP$_4$+OSD$_4$-2 to decode surface, toric, and XZZX codes, achieving thresholds of roughly 17.5\%--17.7\%.
The threshold analysis for the surface codes is  shown in Fig.~\ref{fig:eps_surface},
where we collect at least 10,000 logical error events for each point on the plot.
By using the scaling ansatz method \cite{WHP03}, we obtain a threshold of 17.68\% for BP$_4$+OSD$_4$-2 on surface codes.
This result  improves upon AMBP$_4$, as illustrated in Fig.~\ref{fig:surface}.

To compare our decoder with other known decoders, we construct Table~\ref{tbl:thld}. In \cite{RWBC20}, they use BP$_2$+pOSD$_2$-(2,60) to achieve a threshold of $p=9.9\%$ on the toric codes over bit-flip errors. Assuming that this decoder can correct all $X$ and $Z$ errors of weight $\le np$, we can assume that all Pauli errors of weight $\le np$ can be corrected. In the depolarizing channel, a decoder with a threshold of $\epsilon$ such that $p=2\epsilon/3$ can achieve this (as described in \cite{MMM04}). Therefore, we suppose that BP$_2$+pOSD$_2$-(2,60) has a threshold of $\epsilon=9.9\%\times 3/2 = 14.85\%$  over   depolarizing errors.

Finally, AMBP$_4$ is capable of decoding various families of topological codes. However, after conducting further comparisons, we observe that BP$_4$+OSD$_4$-2 outperforms AMBP$_4$ for various code families. These results are summarized in Table~\ref{tbl:AvsO}, indicating that there is still potential for improving BP algorithms.

\section{Conclusion and Discussions} \label{sec:conclu}

BP-OSD schemes have proven to be highly effective in decoding various quantum codes. We proposed OSD algorithms based quaternary reliability sorting. Using (M)BP$_4$+OSD$_4$ based on quaternary reliability sorting retains the $X/Z$ correlations and achieves superior performance for both CSS and non-CSS codes.

We focused on \mbox{BP$_4$+OSD$_4$-2} in our investigations. However, BP$_4$+OSD$_4$-0 has demonstrated excellent performance for most codes. This is aligned with the observation in \cite{PK19}.

We note that Def.~\ref{def:reliability order} can be adapted to use alternative metrics such as entropy $h(q_i) = -\sum_{B\in\{I,X,Y,Z\}} q_i^B \log q_i^B$ or maximum $\displaystyle \max_{B\in\{I,X,Y,Z\}} q_i^B$. However, using $\ell_i$ from the original definition yielded the best results.

Our BP-OSD scheme can also be extended for data and syndrome errors in the phenomenological noise model \cite{QVRCT21,HB22}. This can be achieved using the technique proposed in \cite{ALB20,KCL21}.

\bibliographystyle{IEEEtran}
\bibliography{myBib_v2.bib}

\end{document}